\documentclass[prb,twocolumn,preprintnumbers,amsmath,amssymb]{revtex4}

\usepackage{graphicx}
\usepackage{dcolumn}
\usepackage{bm}
\usepackage{mathrsfs}
\usepackage{appendix}
\usepackage{bbm}
\usepackage{color}

\def\be{\begin{equation}}
\def\ee{\end{equation}}
\def\bd{\begin{displaymath}}
\def\ed{\end{displaymath}}
\def\-{\phantom{-}}

\def\Jii{J_2/\vert J_1\vert }
\def\Jiii{J_3/\vert J_1\vert }

\begin{document}
%
%

\title{Critical Anisotropies of a Geometrically-Frustrated Triangular-Lattice Antiferromagnet}

\author{M. Swanson$^{1,2}$, J.T. Haraldsen$^{1}$, and R.S. Fishman$^{1}$ \\
\textit{\small{$^1$Materials Science and Technology Division, Oak Ridge National Laboratory, Oak Ridge, Tennessee 37831, USA}} \\
\textit{\small{$^2$North Dakota State University, Fargo, North Dakota 58105, USA}}}

\date{\today}

\begin{abstract}

This work examines the critical anisotropy required for the local stability of the collinear ground states of a geometrically-frustrated 
triangular-lattice antiferromagnet (TLA).  Using a Holstein-Primakoff expansion, we calculate the spin-wave frequencies for the 
1, 2, 3, 4, and 8-sublattice (SL) ground states of a TLA with up to third neighbor interactions.  Local stability requires 
that all spin-wave frequencies are real and positive.  The 2, 4, and 8-SL phases break up into several regions where the 
critical anisotropy is a different function of the exchange parameters.  We find that the critical anisotropy is a continuous function 
everywhere except across the 2-SL/3-SL and 3-SL/4-SL phase boundaries, where the 3-SL phase has the higher critical anisotropy. 

\end{abstract}

\maketitle

\begin{figure}
\includegraphics[width=2.75in]{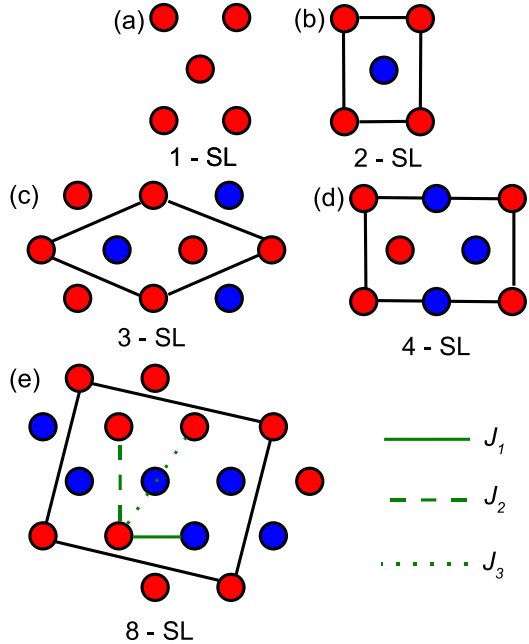}
\caption{(Color online) The 1, 2, 3, 4, and 8-SL phases for the ground states of the geometrically-frustrated TLA. The solid black lines denote the magnetic
unit cell of each phase.  Up and down spins are designated by red and blue circles, respectively.  }
\label{sublattices}
\end{figure}

{\it Introduction}. Geometrically-frustrated systems exhibit many novel characteristics including non-collinear ground states and multiferroic properties\cite{die:04}.  One of the best realizations of a geometrically-frustrated triangular-lattice
antiferromagnet (TLA) is CuFeO$_2$, which contains stacked hexagonal planes of spin-5/2 Fe$^{3+}$ ions.
Accompanied by a phase transition from a collinear 4-sublattice (SL) ground state to a non-collinear phase \cite{mit:91,mek:93,pet:05,ter:06}, 
CuFeO$_2$ exhibits multiferroic properties above a critical magnetic field or above a critical concentration of non-magnetic
Al$^{3+}$ impurities, which substitute for the Fe$^{3+}$ ions \cite{ter:05, sek:07}.  Inelastic neutron-scattering 
experiments \cite{ter:04, kim:06, ye:07} on CuFeO$_2$ have reported
a spin-wave (SW) gap of about 0.9 meV, which decreases with Al doping and may vanish \cite{ter:07} upon the appearance of 
multiferroic behavior.  Similar behavior is produced in a model TLA as the anisotropy is reduced \cite{fis:08} and spin fluctuations 
about the 4-SL collinear phase become stronger.  In this paper, we evaluate the critical anisotropies required for the 
local stability of the collinear magnetic phases in a model TLA with up to third nearest neighbors.  As shown elsewhere \cite{har:08}, 
the wave-vector of the dominant SW instabilities of a collinear phase coincide with the dominant wave-vector of the noncollinear 
phase that appears with decreasing anisotropy.  Therefore, an analysis of the critical anisotropies and wave-vectors of a
frustrated TLA can provide useful information about the non-collinear phases that appear at small anisotropy.

\par
The collinear ground states of a TLA with strong anisotropy were first obtained by Takagi and Makata\cite{tak:95}, who examined an 
Ising model with interactions up to third nearest neighbors.  The ground-state phase diagram consists of the five phases sketched in 
Fig. \ref{sublattices}, where the energies of these five states are given in Table \ref{Energies}.  Using a Holstein-Primakoff (HP) expansion, we have calculated the SW frequencies and critical 
anisotropies for each of these phases.

The Hamiltonian for a TLA is given by
\be
H = -\frac{1}{2}\sum_{i \neq j} J_{ij} \mathbf{{S}}_i \cdot \mathbf{{S}}_j - D \sum_i \mathbf{{S}}_{iz}^2,
\label{genH}
\ee
where $\mathbf{S}_i$ is the local moment on site $i$, $J_{ij}$ is the interaction between sites $i$ and $j$, and $D$ is the single-ion anisotropy.  
Employing a HP transformation, the spin operators are given by $S_{iz} = S -a_{i}^{\dag} a_{i}$, $S_{i+} = \sqrt{2S}a_{i}$, and
$S_{i-} = \sqrt{2S}a_{i}^{\dag}$.  Expanded about the classical limit in powers of $1/\sqrt{S}$, the Hamiltonian can be written as 
$H= E + H_{1} + H_{2}+ \dots$.  The first-order term $H_1$ vanishes when the spin configuration minimizes the energy $E$.  
The second-order term $H_2$ provides the dynamics of non-interacting SWs.  Higher-order terms $H_{n>2}$ reflect the 
interactions between SWs.  They are unimportant at low temperature and for large $1/S$.  Similar to Takagi and Makata, 
we consider nearest neighbor $J_1$, next-nearest neighbor $J_2$, and next-next-nearest-neighbor $J_3$ 
exchange interactions, as sketched in Fig. \ref{sublattices}.  

To determine the SW frequencies $\omega_{\mathbf{k}}$, we solve the equation-of-motion for the vectors $\mathbf{v_{k}} = [a_{\mathbf{k}}^{(1)},
a_{\mathbf{k}}^{(1)\dag},a_{\mathbf{k}}^{(2)}, a_{\mathbf{k}}^{(2)\dag},...]$, which may be written in terms of the 
$2N \times 2N$ matrix $ \underline{M}(\mathbf{k})$ as 
$id\mathbf{v_k}/dt = -\big[ \underline{H}_{2},\mathbf{v_k}\big] =  \underline{M}(\mathbf{k}) \mathbf{v_k}$,
where $N$ is the number of spin sites in the unit cell.  The SW frequencies are
then determined from the condition Det[$  \underline{M}(\mathbf{k}) - \omega_{\mathbf{k}} \underline{I}$] = 0.

Two conditions are required for the local stability of any magnetic phase:  all SW frequencies must be real and positive and all 
SW weights must be positive.  The SW weights $W_{\mathbf{k}}^{(s)}$ are coefficients of the spin-spin correlation function:
\be
\begin{array}{c}
\displaystyle S(\mathbf{k},\omega) =  \frac{1}{ N} \int dt~ e^{-i \omega t}
\sum_{i,j} e^{i \mathbf{k\cdot d_{ij}}}\Big\{ \big< \mathbf{S}_i^{+}\mathbf{S}_j^{-} (t)\big> \\ \\ \displaystyle
 +\big< \mathbf{S}_i^{-}\mathbf{S}_j^{+} (t)\big> \Big\}= \sum_{s} W_{\mathbf{k}}^{(s)}\delta(\omega-\omega_{\mathbf{k}}^{(s)}),
\end{array}
\label{weights}
\ee
where $s$ denotes a branch of the SW spectrum and $\mathbf{d}_{ij}$ is defined as the vector pointing from site $i$ to site $j$.
The weights $W_{\mathbf{k}}^{(s)}$ were evaluated within the HP formalism by solving the equations-of-motion for coupled spin
Green's functions \cite{har:09,fis:08b}.  In zero field, the condition that the SW weights are positive for all $\mathbf{k}$ is equivalent to the 
condition that all SW frequencies are positive.

\begin{table}
\caption{\textbf{Classical Energies and Critical Anisotropies for TLA Sublattices}}
\begin{ruledtabular}
\begin{tabular}{llc}
 \textbf{SL} &  \textbf{Energy}  & $\textbf{D}_c$ \\
\hline \\
1-SL & $\displaystyle \frac{E^{(1)}}{NS^2} = -3J_1-3J_2-3J_3-D$ & $D_c^{(1)}$ = 0 \\ \\
\hline \\
2-SL & $\displaystyle \frac{E^{(2)}}{NS^2} = J_1+J_2-3J_3-D$ & $D_c^{(2\rm{I})}$ (Eq. (\ref{Dc2I}))  \\
 &  & $D_c^{(2\rm{II})}$ (Eq. (\ref{Dc2II})) \\
 &  & $D_c^{(2\rm{III})}$ (Eq. (\ref{Dc2III})) \\
& & $D_c^{(2\rm{IV})}$ = 0  \\ \\
\hline \\
3-SL & $\displaystyle \frac{E^{(3)}}{NS^2} = J_1-3J_2+J_3-D$ & $D_c^{(3)}$ (Eq. (\ref{Dc3}))  \\ \\
\hline \\
4-SL & $\displaystyle \frac{E^{(4)}}{NS^2} = J_1-J_2+J_3-D$ & $D_c^{(4\rm{I})}$ (Eq. (\ref{Dc4I}))  \\
&  & $D_c^{(4\rm{II})}$ (Eq. (\ref{Dc4II}))  \\ \\
\hline \\
8-SL & $\displaystyle \frac{E^{(8)}}{NS^2} = J_2+J_3-D$ & $D_c^{(8\rm{I})}$ (Eq. (\ref{Dc8I}))  \\
&  & $D_c^{(8\rm{II})}$ (Eq. (\ref{Dc8II}))  \\ \\
\end{tabular}
\end{ruledtabular}
\label{Energies}
\end{table}

We obtained analytic expressions for the SW frequencies for all phases shown in Fig. \ref{sublattices} with 
the exception of the 8-SL phase, which was solved numerically.  Analysis of the SW frequencies yields the critical anisotropy $D_c$ 
and the critical wave-vectors $\mathbf{k}$ where the SW frequencies vanish.  To simplify the following discussion, the SW and anisotropy 
coefficients are provided in the appendix.

\begin{figure}
\includegraphics[width= 3.75in]{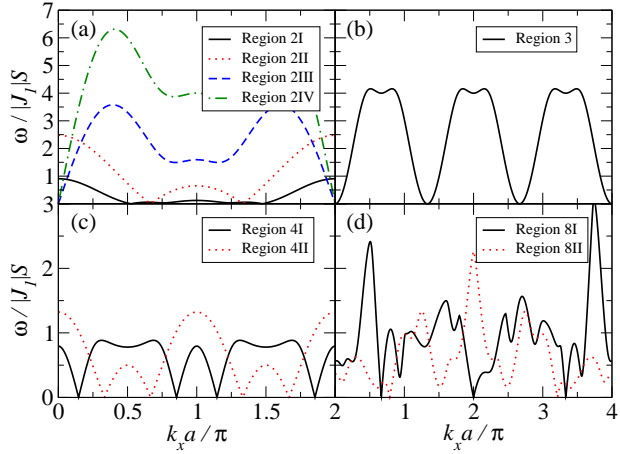}
\caption{(Color online) SW frequencies at the critical anisotropy for 2 (a), 3 (b), 4 (c), and 8-SL (d) phases (interaction parameters given in the text).  
All SW instabilities occur for $k_y a =0$ except in regions 2III and 8I, where they occur for $k_ya$ = 0.186$\pi$ and 0.382$\pi$, respectively.}
\label{spinwaves}
\end{figure}

\begin{figure}
\includegraphics[width= 0.48 \textwidth]{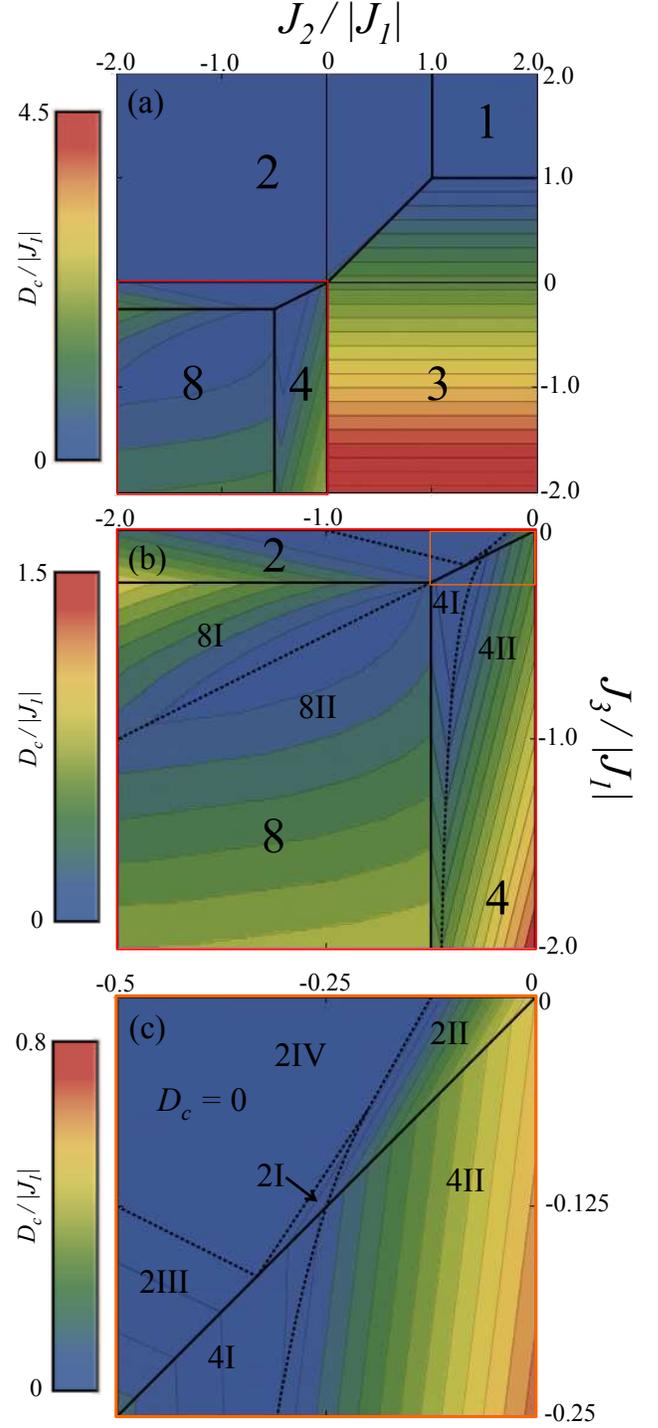}
\caption{(Color online) Critical anisotropies for the TLA ground states.  The spacing between contours is 0.2 (a), 0.1 (b), and 0.05 (c).  
Numbers designate the stable phase and Roman numerals designate regions where the behavior of the critical anisotropy
is distinct.  Solid lines denote boundaries between phases and dashed lines denote boundaries between regions.  
$D_c$ is continuous across each phase boundary except for the 2-SL/3-SL and the 3-SL/4-SL phase boundaries;  in both cases,
$D_c$ is higher for the 3-SL phase.}
\label{contour}
\end{figure}

{\it1 - Sublattice}. The 1-SL phase (Fig. \ref{sublattices}(a)) is a ferromagnet with SW frequencies 
\be
\omega_{\mathbf{k}}^{(1)} = 2 S \Big(D + A_{1\mathbf{k}}\Big).
\ee 
Since the 1-SL phase is locally stable for any positive value of the anisotropy, $D_c = 0$. The SW intensity $W_{\mathbf{k}}^{(1)}$ is constant throughout $\mathbf{k}$ for all interactions.

{\it2 - Sublattice}. For the 2-SL phase (shown in Fig. \ref{sublattices}(b)), the SW frequencies are given by
\be 
\omega_{\mathbf{k}}^{(2)} = 2S \sqrt{A_{2\mathbf{k}}^2 - A_{3\mathbf{k}}^2}.
\label{omega2}
\ee
The SW weights for the 2-SL phase are
\be
W_{\mathbf{k}}^{\rm{(2)}} = \sqrt{\frac{A_{2\mathbf{k}}+A_{3\mathbf{k}}}{A_{2\mathbf{k}}-A_{3\mathbf{k}}}}.
\ee
From Eq. (\ref{omega2}), the condition for the local stability of a 2-SL phase is $A_{2\mathbf{k}}^2 - A_{3\mathbf{k}}^2 > 0$.  At $D_c$, 
$A_{2\mathbf{k}}^2 = A_{3\mathbf{k}}^2$.  This condition is satisfied when $D_c = 0$ in most of the 2-SL phase.  But approaching the 
3, 4, and 8-SL phase boundaries, nonzero anisotropy is required for local stability.  As shown in Fig. \ref{contour}(a), the critical anisotropy is 
continuous across the 4-SL and 8-SL boundaries, but is discontinuous across the 3-SL boundary. 

Upon closer examination (Fig. \ref{contour}(c)), we find that $D_c$ depends differently on the exchange parameters in the three regions
designated by Roman numerals.  In region 2I (bounded by $ J_3 = J_2/2$, $J_3 = (9 J_2 - J_1)/12$, and $J_3 = J_2^2/(J_1 -2 J_2)$),
\be
\begin{array}{l}
D_c^{(2\rm{I})} = \dfrac{1}{(4 J_3)^3} \Big\{-272 J_3^4 + 64 J_3^3 J_2 + 48 J_3^3 J_1 \\ \\
       \hspace{1 cm}   + 72 J_3^2 J_2^2 - 48 J_3^2 J_2 J_1 - 8 J_3^2 J_1^2 \\ \\
       \hspace{1 cm} + 36 J_3 J_2^2 J_1 - 27 J_2^4  - (2 J_3 - J_2) C^3 \Big\},
\end{array}
\label{Dc2I}
\ee
where
\be
C = \sqrt{(2 J_3 + 3 J_2)^2 - 8 J_3 J_1}
\label{kc2III}
\ee
In region 2II (bounded by  $J_3 = J_2/2$, $J_3 = J_2$, $J_3 = (8 J_2 - J_1)/9$, and $J_3 = J_2^2/(J_1 -2 J_2)$),
\be
D_c^{(2\rm{II})} = 4 J_2 - \frac{9}{2} J_3 - \frac{1}{2} J_1.
\label{Dc2II}
\ee
Finally, in region 2III (bounded by $J_3 = J_2/2$, $J_3 = J_1/4$, and $J_3 = (J_1 - J_2)/4$),
\be
D_c^{(2\rm{III})} = - \frac{(4 J_3 + J_2 - J_1)^2}{2(J_2 + 2 J_3)}.
\label{Dc2III}
\ee
The region with no critical anisotropy $D_c^{(2\rm{IV})}$ is bounded by $J_2$ = 0, $J_3 = (8 J_2 - J_1)/9$, and $J_3 = (J_1 - J_2)/4$ as shown in Figs. \ref{contour}(b) and (c).

The critical wave-vectors, $\mathbf{k}$ for the SW instabilities in region 2I are:
\be
\begin{array}{l}
k_x^{(2\rm{I, a})} a = 2  \arccos\left\{\dfrac{3 J_2 - 2 J_3 - C}{8 J_3} \right\}, \\ \\ 
k_y^{(2\rm{I, a})} a = 0.
\end{array}
\label{kc2Ia}
\ee
Two other instabilities $\mathbf{k}^{(2\rm{I, b})}$ and $\mathbf{k}^{(2\rm{I, c})}$ are related to $\mathbf{k}^{(2\rm{I, a})}$ by $\pm \pi /3$ rotations and 
can be considered ``twins" of the ${\bf k}^{(2\rm{I,a})}$ instabilities.  All three instabilities occur at the same critical anisotropy $D_c^{(2\rm{I})}$.  
For regions 2II and III, the SW instabilities occur at
\be 
\begin{array}{l}
k_x^{(2\rm{II})} a = \pi \pm \pi/3, \\ \\
k_y^{(2\rm{II})} a = 0,
\end{array}
\label{kc2II}
\ee
and
\be
\begin{array}{l}
k_x^{(2\rm{III})} a = 0, \\ \\
k_y^{(2\rm{III})} a = \dfrac{2}{\sqrt{3}} \arccos \left\{\dfrac{J_2 + J_1}{2(J_2 + 2 J_3)}\right\},
\end{array}
\label{kc2III}
\ee
along the $k_x$ and $k_y$ axis, respectively. 

Figure \ref{spinwaves}(a) shows three representative SWs for all 2-SL regions.  The interaction parameters for region 2I are
$J_2/\vert J_1\vert  = -0.25$, $J_3/\vert J_1\vert  = -0.12$, and $D_c/\vert J_1\vert = 0.04$.
For region 2II, $J_2/\vert J_1\vert  = -0.10$, $J_3/\vert J_1\vert  = -0.05$, and $D_c/\vert J_1\vert  = 0.325$.  For region 2III,
$J_2/\vert J_1\vert  = -0.75$, $J_3/\vert J_1\vert  = -0.125$, and $D_c/\vert J_1\vert  = 0.031$. Finally, for region IV, the interaction parameters are $J_2/\vert J_1\vert  = -1.0$, $J_3/\vert J_1\vert  = -0.125$, and $D_c/\vert J_1\vert  = 0.0$.
Regions I, II and IV were evaluated with $k_ya = 0$, while region III was evaluated at $k_ya = 0.186\pi$ as explained above.

In Figs. \ref{contour}(b) and (c), we examine the critical anisotropy of the 2-SL along the $J_3/\vert J_1\vert $ = 0 axis.   The critical anisotropy 
vanishes for $-1 < J_2/\vert J_1\vert  < -1/8$ but is nonzero outside this region.  Therefore, non-collinear phases should 
appear for $J_2/\vert J_1\vert < -1$ and $J_2/\vert J_1\vert > -1/8$ when $D < D_c$.  This agrees with Jolicoeur $et~al.$ \cite{jol:90}, 
who studied a TLA with nearest and next-nearest neighbor exchange interactions and $D = 0$.  
They obtain a N\'eel state up to $J_2/\vert J_1\vert  = -1/8$ and an incommensurate spiral for
$J_2/\vert J_1\vert  < -1$.  Similar results have been obtained on square lattices \cite{dag:89,cha:88}.

\begin{figure}
\includegraphics[width= 0.5 \textwidth]{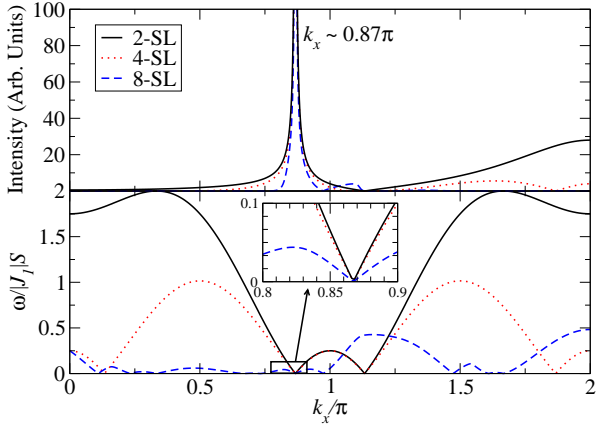}
\caption{(Color online) SW frequencies and SF intensities for the 2, 4, and 8-SL phases at the critical anisotropy with 
$J_2/\vert J_1\vert  = -0.5$, $J_3/\vert J_1\vert  = -0.25$, and $D/\vert J_1\vert  = 0.125$.}
\label{248-I}
\end{figure}

{\it3 - Sublattice}.  For the 3-SL phase (shown in Fig. \ref{sublattices}(c)), the SW frequencies are
\be
\omega_{\mathbf{k}}^{(3)} = 6S \,\sqrt {R_{1\mathbf{k}}}\cos \left(\theta/3 + 2m/3\pi \right) +R_{2\mathbf{k}}/3,
\ee
where $m$ is an integer (0,1,2) distinguishing the three separate SW dispersion relations and
\be
\theta = \arccos \left\{\frac{2R_{2\mathbf{k}}^3-9R_{2\mathbf{k}}R_{3\mathbf{k}}-27R_{4\mathbf{k}}}{1458{R_{1\mathbf{k}}^{3/2}}}\right\}.
\ee
The critical anisotropy of the 3-SL phase is independent of $J_2$ and given by
\be
D_c^{(3)} = - \frac{3}{2}(J_1 + J_3).
\label{Dc3}
\ee
Notice find that $D_c^{(3)} = 0$ along the 3-SL/1-SL boundary.
Again, $D_c$ is discontinuous along the 2-SL/3-SL and 3-SL/4-SL boundaries:  
the anisotropy required for the local stability of the 3-SL phase is three times the critical anisotropy of the 2 or 4-SL phases.  
As discussed further below, the discontinuities at the 2-SL/3-SL and 3-SL/4-SL phase boundaries are related to the distinction 
between the conditions for global and local stability.

In Fig. \ref{spinwaves}(b), we plot a SW dispersion in the 3-SL phase 
with interaction parameters $J_2/\vert J_1\vert = 0.5$, $J_3/\vert J_1\vert  = -0.5$, and $D_c/\vert J_1\vert = 2.25$.  
Since the 3-SL phase has a net moment, the SW frequencies are quadratic functions of $\mathbf{k}$ near the 
instability wave-vectors. 

\begin{figure*}
\includegraphics[width= 0.98 \textwidth]{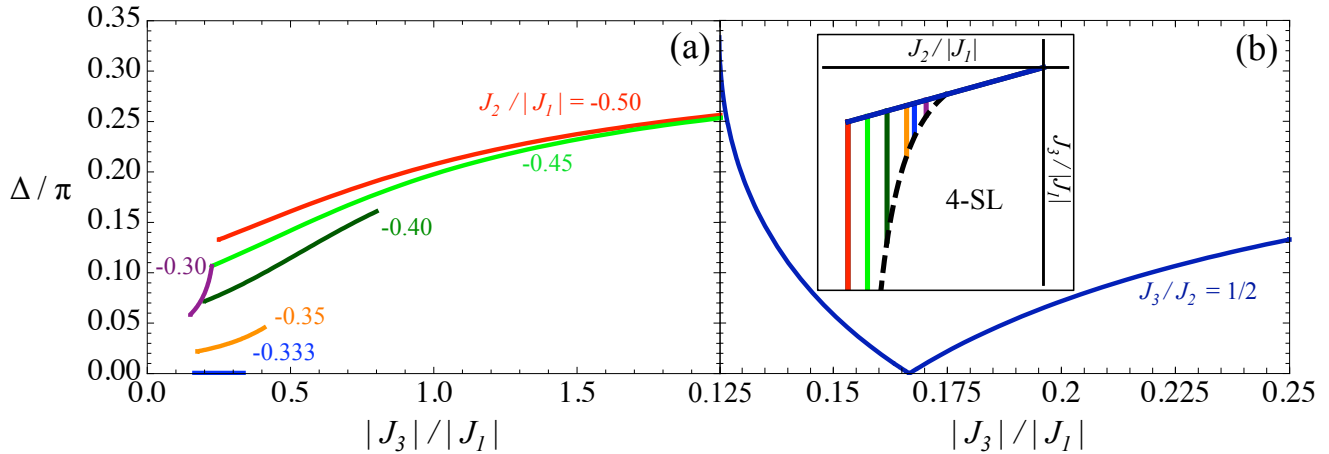}
\caption{(Color online) (a)  Location of SW instability $\Delta = a \vert k_x - \pi\vert $ along $k_ya = 0$ in region 4I for 
fixed values of $\Jii$.  As $\vert J_3\vert /\vert J_1\vert $ increases along $\Jii = -0.5$, $\Delta$ asymptotically approaches $\pi/3$.  
(b)  Plot of $\Delta$ in region 4I along the 2-SL/4-SL boundary $J_3/J_2 = 1/2$.  The cusp in $\Delta $ occurs at $J_2/\vert J_1\vert  = -1/3$ 
where the SW instability occurs at $\pi$.}
\label{kx}
\end{figure*}

{\it4 - Sublattice}. The SW frequencies for the 4-SL phase (shown in Fig. \ref{sublattices}(d)) were evaluated
in Ref. [\onlinecite{fis:08}] and are given by
\be
\begin{array}{c}
\omega_{\mathbf{k}}^{(4)} = 2S\Big(A_{6\bf{k}}^2 - A_{7\mathbf{k}}^2 \pm \Big[ \Big(F_{2\mathbf{k}}^2 - F_{2\mathbf{k}}^{*2}\Big)^2 \\ \\ + 4\vert A_{6\mathbf{k}}F_{2\mathbf{k}} - A_{7\mathbf{k}}F_{2\mathbf{k}}^*\vert ^2\Big]^{1/2}\Big)^{1/2}.
\end{array}
\ee
The SW weights of the 4-SL phase are
\be
\begin{array}{c}
W_{\mathbf{k}}^{(\rm{4})} =  \Big[R_{5\mathbf{k}}(A_{7\mathbf{k}}-A_{6\mathbf{k}})
+(F_{2\mathbf{k}}+F_{2\mathbf{k}}^*)(A_{6\mathbf{k}}-A_{7\mathbf{k}})^2 \\ \\
+(F_{2\mathbf{k}}-F_{2\mathbf{k}}^*)^2(F_{2\mathbf{k}}+F_{2\mathbf{k}}^*-A_{6\mathbf{k}}-A_{7\mathbf{k}}) \Big] \\ \\
\times \Big[R_{5\mathbf{k}}\sqrt{A_{6\mathbf{k}}^2-A_{7\mathbf{k}}^2-R_{5\mathbf{k}}}\Bigr]^{-1}.
\end{array}
\ee
As for the 2-SL phase, the critical anisotropy $D_c$ for the 4-SL phase depends differently on the interaction parameters
in two regions, again denoted by Roman numerals I and II 
(Fig. \ref{contour}(b)).  In region 4I (bounded by $J_3 = J_2/2$, $J_2 = J_1/2$, and $J_3 = J_2^2/(J_1 -2 J_2)$),
\be
\begin{array}{l}
D_c^{(4\rm{I})} = \dfrac{1}{(4 J_3)^3} \Big\{-16 J_3^4 - 64 J_3^3 J_2 + 48 J_3^3 J_1   \\ \\
       \hspace{1 cm}   + 72 J_2^2 J_3^2 - 8 J_3^2 J_1^2 - 48 J_3^2 J_2 J_1   \\ \\ 
       \hspace{1 cm}   + 36 J_3 J_2^2 J_1 - 27 J_2^4 + (2 J_3 - J_2) C^3 \Big\}
\end{array}
\label{Dc4I}
\ee
and in region 4II (bounded by $J_3 = J_2/2$, $J_2 = 0$, and $J_3 = J_2^2/(J_1 -2 J_2)$),
\be
D_c^{(4\rm{II})} = 2 J_2 - \frac{1}{2} J_3 - \frac{1}{2} J_1.
\label{Dc4II}
\ee
The critical wave-vectors for the 4-SL phase are the same as those in the respective region of the 2-SL phase, including the multiple instabilities in region 2I: 
$\mathbf{k}^{(4\rm{I,a})} = \mathbf{k}^{(2\rm{I,a})}$, $\mathbf{k}^{(4\rm{I,b})} = \mathbf{k}^{(2\rm{I,b})}$,  and $\mathbf{k}^{(4\rm{II})} = \mathbf{k}^{(2\rm{II})}$. 
Figure \ref{spinwaves}(c) shows two representative SWs for regions 4I and 4II with $k_ya =0$. The interactions parameters for region 4I are 
$J_2/\vert J_1\vert = -0.439$, $J_3/\vert J_1\vert = -0.570$, and $D_c/\vert J_1\vert  = 0.105$.  For region 4II, 
they are $J_2/\vert J_1\vert = -0.25$, $J_3/\vert J_1\vert = -0.5$, and $D_c/\vert J_1\vert = 0.25$.

Figure \ref{248-I} shows the SW frequencies and 
SF intensities for the 2, 4, and  8-SL phases at the triple point of the phase diagram. The intensities for the 8-SL phase were determined numerically. 
As shown in Fig. \ref{248-I}, each phase becomes unstable at $D/\vert J_1\vert = 0.125$, where the SW intensity for each phase peaks at the same wave-vector. 
This wave-vector corresponds to the ordering wave-vector of the non-collinear phase \cite{har:08} that appears at small anisotropy. 
Because the 2, 4, and 8-SL phases all have zero net moment,  their SW frequencies are linear functions of $\mathbf{k}$ around the wave-vectors 
of the instabilities.

The 4-SL phase is of particular interest since it is the known ground state \cite{mit:91} of CuFeO$_2$.  Fits of the experimental SW frequencies \cite{ye:07, fis:08b} 
of  CuFeO$_2$ have determined the ratios of exchange parameters $\Jii \approx -0.44$ and $\Jiii \approx -0.57$, which lies within region 4I.  
Consequently, we have studied the SW frequencies of the 4-SL phase more closely.  
Figure \ref{kx} shows the behavior of $k_x^{(4\rm{I},a)}$ along various cuts through region 4I of phase space.  Since the SW frequencies are symmetric about the 
midpoint of the Brillouin zone $a\pi$, we consider the quantity $\Delta \equiv a | k_x - \pi |$.  As $\Jiii$ increases in region 4I, $\Delta$ 
asymptotically approaches $\pi/3$,  which is the constant value of $\Delta$ in region 4II.  For small values of $\Jiii$, the wave-vector instabilities approach 
$\pi$ as $\Jii$ increases, equal $\pi$ for $\Jii = -1/3$, and then move away from $\pi$ as $\Jii$ approaches zero; this behavior is shown along the 
2-SL/4-SL boundary in Fig. \ref{kx}(b).

{\it8 - Sublattice}. For the 8-SL phase (shown in Fig. \ref{sublattices}(e)), we have determined SW dispersion relations numerically.  
The critical anisotropy values for this phase are shown in Fig. \ref{contour}(a).  
Notice that $D_c$ has a cusp dividing the phase into regions 8I and 8II (Fig. \ref{contour}(b)), separated by $J_3=J_2/2$.  
Looking more closely at the numerical results, 
the critical anisotropies in the 8-SL regions are closely related to those of their respective neighbors and are given by
\be
D_c^{(8\rm{I})} = D_c^{(2\rm{III})} + 4 J_3 - J_1,
\label{Dc8I}
\ee
\be
D_c^{(8\rm{II})} = D_c^{(4\rm{I})} + 2 J_2 - J_1,
\label{Dc8II}
\ee
which clearly show that the critical anisotropies are continuous across the phase boundaries.  In region 8II, the wave-vector instabilities occur 
for $k_y = 0$ (as in region 4I);  in region 8I, the wave-vector instabilities occur for non-zero $k_y$ (as in region 2III).  Figure \ref{spinwaves}(d) 
shows two representative SWs for regions 8I and 8II. The interactions parameters for region 8I are $J_2/\vert J_1\vert  = -1.5$, 
$J_3/\vert J_1\vert = -0.50$, and $D_c/\vert J_\vert  = 0.25$.  For region 8II, they are $J_2/\vert J_1\vert  = -0.75$, $J_3/\vert J_1\vert  = -0.50$, 
and $D_c/\vert J_1\vert  = 0.62$.  Whereas $k_ya=0$ for region 8II, $k_ya = 0.382\pi$ for region 8I as explained above.

To better understand the discontinuities along the 2-SL/3-SL and 3-SL/4-SL phase boundaries, we consider the relationship between 
local and global stability.  Our SW calculations only guarantee the local stability of each collinear phase.  But even when a 
phase is locally stable, it can still be globally unstable to a lower-energy spin configuration.  Hence, the critical anisotropy $\tilde{D}_c$
for global stability must be greater than or equal to the critical anisotropy $D_c$ for local stability.
Unlike $D_c$, $\tilde{D}_c$ must also be a continuous function of $J_1$, $J_2$, and $J_3$.  So when $D_c$ is discontinuous, 
the phase with the lower critical anisotropy cannot be globally stable.  Since the 3-SL has a higher critical anisotropy along the 2 and 4-SL boundaries, 
the 2-SL and 4-SL phases cannot be globally stable along those boundaries when $D_c^{(2{\rm II})} <  D < D_c^{(3)}$ or $D_c^{(4{\rm II})} <  D < D_c^{(3)}$.  
Therefore, our results for the local stability of the collinear phases also has implications for the global stability of those phases.

{\it Conclusion}. We have examined the critical anisotropy for a geometrically-frustrated TLA.  Based on the Takagi-Makata phase diagram, 
we calculated the SW frequencies for all five phases.  Imposing the two conditions for local stability, we obtained the critical anisotropies 
and wave-vector instabilities for all phases as functions of the exchange interactions.  Surprisingly, these results are highly dependent 
on the longer-range exchange interactions and most phases break into several regions where the anisotropy has a distinct 
dependence on the exchange parameters.   As discussed for the 2-SL and 4-SL phases, the critical anisotropies and wave-vectors 
for the local stability of the collinear phases provides useful information about the non-collinear phases that appear at small anisotropy.  
We have also shown that the discontinuity of the critical anisotropy at the 2-SL/3-SL and 3-SL/4-SL phase boundaries has implications 
for the global stability of the 2-SL and 4-SL phases with the smaller critical anisotropies.

We would like to acknowledge helpful conversations with Gonzalo Alvarez.  This research was sponsored by the Laboratory Directed Research 
and Development Program of Oak Ridge National Laboratory, managed by UT-Battelle, LLC for the U.S. Department of Energy under 
contract No. DEAC05-00OR22725 and by the Division of Materials Science.  
We would also like to acknowledge the DOE SULI program for support during this research.

\appendix

\section{Spin-wave and Anisotropy Coefficients}

This Appendix provides the coefficients that enter the SW frequencies and weights for each phase. The coefficients
for the 1-SL or ferromagnetic phase are
\be
\begin{array}{l}
A_{1\mathbf{k}} = 3(J_1+J_2+J_3) \\ \\
\hspace{1cm} -J_1 \Big(  \cos{(\mathbf{k} \cdot \mathbf{d}_1)} + \cos{(\mathbf{k} \cdot \mathbf{d}_2)} + \cos{(\mathbf{k} \cdot \mathbf{d}_3)} \Big) \\ \\
\hspace{1cm} -J_2 \Big( \cos{(\mathbf{k} \cdot \mathbf{d}_4)} + \cos{(\mathbf{k} \cdot \mathbf{d}_5)} + \cos{(\mathbf{k} \cdot \mathbf{d}_6)} \Big)  \\ \\
\hspace{1cm} -J_3 \Big( \cos{(2 \mathbf{k} \cdot \mathbf{d}_1)} + \cos{(2 \mathbf{k} \cdot \mathbf{d}_2)}+ \cos{(2 \mathbf{k} \cdot \mathbf{d}_3)} \Big),
\end{array}
\label{1-SLCoef}
\ee
where $\mathbf{d}_1=\,{\it a \mathbf{x}}$, $\mathbf{d}_2=1/2\,{\it a \mathbf{x}}+\sqrt {3}/2\,{\it a \mathbf{y}}\,$, 
$\mathbf{d}_3=-1/2\,{\it a \mathbf{x}}+\sqrt {3}/2\,{\it a \mathbf{y}}\,$, $\mathbf{d}_4=3/2\,{\it a \mathbf{x}}+\sqrt {3}/2\,{\it a \mathbf{y}}\,$, 
$\mathbf{d}_5=\sqrt {3}\,{\it a \mathbf{y}}\,$, and $\mathbf{d}_6=-3/2\,{\it a \mathbf{x}}+\sqrt {3}/2\,{\it a \mathbf{y}}\,$. 

The 2-SL phase coefficients are
\be
\begin{array}{l}
A_{2\mathbf{k}} =D +3J_3 \\ \\ \hspace{1cm} -  J_1 \Big(\cos{(\mathbf{k} \cdot \mathbf{d}_1)} + 1 \Big) - J_2 \Big(\cos{(\mathbf{k} \cdot \mathbf{d}_5)} + 1 \Big) \\ \\ \hspace{1cm}
- J_3 \Big( \cos{(2 \mathbf{k} \cdot \mathbf{d}_1)} + \cos{(2 \mathbf{k} \cdot \mathbf{d}_2)} + \cos{(2 \mathbf{k} \cdot \mathbf{d}_3)} \Big),
\end{array}
\ee
\be
\begin{array}{l}
A_{3\mathbf{k}} = J_1 \Big(\cos{(\mathbf{k} \cdot \mathbf{d}_2)} + \cos{(\mathbf{k} \cdot \mathbf{d}_3)}  \Big) \\ \\
\hspace{1cm} + J_2 \Big(\cos{(\mathbf{k} \cdot \mathbf{d}_4)} + \cos{(\mathbf{k} \cdot \mathbf{d}_6)}  \Big).
\end{array}
\label{2-SLCoef}
\ee

The 3-SL phase coefficients are 
\be
\begin{array}{l}
R_{1\mathbf{k}} = R_{2\mathbf{k}}^2-3R_{3\mathbf{k}},
\end{array}
\ee
\be
\begin{array}{l}
R_{2\mathbf{k}} = 2A_{4\mathbf{k}}+A_{5\mathbf{k}},
\end{array}
\ee
\be
\begin{array}{l}
R_{3\mathbf{k}} = A_{4\mathbf{k}}^2+2A_{4\mathbf{k}}A_{5\mathbf{k}}+\vert F_{1\mathbf{k}}\vert^2, 
\end{array}
\ee
\be
\begin{array}{l}
R_{4\mathbf{k}} = (A_{5\mathbf{k}}-2A_{4\mathbf{k}})\vert F_{1\mathbf{k}}\vert^2 -A_{4\mathbf{k}}^2A_{5\mathbf{k}}   -F_{1\mathbf{k}}^3-F_{1\mathbf{k}}^{*3} ,
\end{array}
\ee

\be
\begin{array}{l}
A_{4\mathbf{k}}= 2D+2J_2(3-\cos(\mathbf{k}\cdot\mathbf{d}_4) \\ \\ \hspace{1cm} -\cos(\mathbf{k}\cdot\mathbf{d}_5)-\cos(\mathbf{k}\cdot\mathbf{d}_6)), 
\end{array}
\ee
\be
\begin{array}{l}
A_{5\mathbf{k}}= 6J_1+6J_3-A_{3\mathbf{k}},
\end{array}
\ee
\be
\begin{array}{l}
F_{1\mathbf{k}} = J_1(e^{-i\mathbf{k}\cdot\mathbf{d}_2}+e^{i\mathbf{k}\cdot\mathbf{d}_1}+e^{i\mathbf{k}\cdot\mathbf{d}_3})
\\ \\ \hspace{1cm} 
+J_3(e^{2i\mathbf{k}\cdot\mathbf{d}_2}+e^{-2i\mathbf{k}\cdot\mathbf{d}_1}+e^{-2i\mathbf{k}\cdot\mathbf{d}_3}).
\end{array}
\label{3-SLCoef}
\ee

As in Ref.(\onlinecite{fis:08}), the 4-SL phase coefficients are 
\be
\begin{array}{l}
R_{5\mathbf{k}} = \Big(F_{2\mathbf{k}}^4+F_{2\mathbf{k}}^{*4} - 2(F_{2\mathbf{k}}^{*2}+2A_{6\mathbf{k}}A_{7\mathbf{k}})F_{2\mathbf{k}}^2 \\ \\ \hspace{1cm}
+4(A_{6\mathbf{k}}^2+A_{7\mathbf{k}}^2)\vert F_{2\mathbf{k}}\vert^2 -4A_{6\mathbf{k}}A_{7\mathbf{k}}F_{2\mathbf{k}}^{*2}\Big)^{1/2},
\end{array}
\ee
\be
\begin{array}{l}
A_{6\mathbf{k}} =  D - J_1 + J_2(1-\cos(\mathbf{k} \cdot \mathbf{d}_5) \\ \\
\hspace{1cm}-J_3(1+\cos(2\mathbf{k} \cdot \mathbf{d}_1)),
\end{array}
\ee
\be
\begin{array}{l}
A_{7\mathbf{k}} = -\cos(\mathbf{k}\cdot\mathbf{d}_1)\Big(J_1 + 2J_3\cos(\sqrt{3}\mathbf{k}\cdot\mathbf{d}_5)\Big),
\end{array}
\ee
\be
\begin{array}{l}
F_{2\mathbf{k}} = -\cos(\mathbf{k} \cdot \mathbf{d_5}/2)  \Big(J_1e^{i\mathbf{k}\cdot\mathbf{d}_1/2} + J_2e^{-3i\mathbf{k} \cdot \mathbf{d}_1/2}\Big).
\end{array}
\label{4-SLCoef}
\ee

\end{document}